%% file: main.tex
\documentclass{llncs}
\usepackage[utf8]{inputenc}
\usepackage{multirow}

\usepackage{amsmath}
\usepackage{amsopn}
\DeclareMathOperator{\cut}{cut}
\DeclareMathOperator{\vol}{vol}

\DeclareMathOperator{\plogp}{plogp}
\newcommand*\rot{\rotatebox{90}}

\usepackage{todonotes}
\usepackage{booktabs}

\usetikzlibrary{positioning,calc,shapes}
\tikzstyle{node}=[circle,inner sep=0.5mm,minimum size=5.25mm,draw = black]
\tikzstyle{blue}=[fill=blue]
\tikzstyle{red}=[fill=red]
\tikzstyle{green}=[fill=green]
\tikzstyle{cyan}=[fill=cyan]
\tikzstyle{magenta}=[fill=magenta]
\tikzstyle{yellow}=[fill=yellow]
\tikzstyle{white}=[fill=white]
\tikzstyle{black}=[fill=black]
\tikzstyle{bright}=[fill=black!14]
\tikzstyle{dark}=[fill=black!28]

\tikzstyle{highlight}=[circle, minimum size=7.5mm, draw = black, dashed]
\tikzstyle{invisible}=[opacity=0]
\tikzstyle{visible}=[color=black, opacity=1]
\tikzstyle{delta_q}=[midway, invisible, font=\tiny]
\tikzstyle{best}=[color=red]

\tikzstyle{item}=[rectangle,inner sep=0.5mm,minimum width=4mm, minimum height=4mm, node distance=4mm,draw = black]
\tikzstyle{func}=[circle,inner sep=0.5mm,minimum width=3mm, minimum height=3mm, draw = black, font=\tiny]

\tikzstyle{label}=[outer sep=0.5mm, minimum size=1.5mm, inner sep=0pt, invisible, circle, draw=black]

\tikzstyle{lop}=[draw,inner sep=0.2em, minimum width=3cm,rectangle]
\tikzstyle{dop}=[lop,ellipse]
\tikzstyle{action}=[lop,chamfered rectangle]
\tikzstyle{code}=[lop,inner sep=0.5em]

\newcounter{nodeCounter}[figure]
\newcommand{\node}[2][]{\refstepcounter{nodeCounter}\ifthenelse{\equal{#1}{}}{}{\label{#1}}#2}

\usepackage{hyperref}
\title{Distributed Graph Clustering using Modularity and Map Equation
\thanks{This work was partially supported by the DFG under grants WA654/19-2 and WA654/22-2. The authors acknowledge support by the state of Baden-Württemberg through bwHPC.}
}

\author{Michael Hamann \and Ben Strasser \and Dorothea Wagner \and Tim Zeitz}
\institute{Institute of Theoretical Informatics, Karlsruhe Institute of Technology, Karlsruhe, Germany;
  \email{michael.hamann@kit.edu}, \email{academia@ben-strasser.net}, \email{dorothea.wagner@kit.edu}, and \email{tim.zeitz@kit.edu}}

\begin{document}

\newcommand{\dslmmod}{DSLM-Mod}
\newcommand{\dslmmodwocont}{DSLM-Mod w/o Cont.}
\newcommand{\dslmmap}{DSLM-Map}

\maketitle

\begin{abstract}
We study large-scale, distributed graph clustering.
Given an undirected graph, our objective is to partition the nodes into disjoint sets called clusters.
A cluster should contain many internal edges while being sparsely connected to other clusters.
In the context of a social network, a cluster could be a group of friends.
Modularity and map equation are established formalizations of this internally-dense-externally-sparse principle.
We present two versions of a simple distributed algorithm to optimize both measures.
They are based on Thrill, a distributed big data processing framework that implements an extended MapReduce model.
The algorithms for the two measures, \dslmmod{} and \dslmmap{}, differ only slightly.
Adapting them for similar quality measures is straight-forward.
We conduct an extensive experimental study on real-world graphs and on synthetic benchmark graphs with up to 68 billion edges.
Our algorithms are fast while detecting clusterings similar to those detected by other sequential, parallel and distributed clustering algorithms.
Compared to the distributed GossipMap algorithm, \dslmmap{} needs less memory, is up to an order of magnitude faster and achieves better quality.
\end{abstract}

\section{Introduction}

Graph clustering is a well researched topic~\cite{f-c-09,fh-ca-16} and has many applications, such as community detection in social networks where users can be modeled as nodes and friendships as edges between them.
These graphs can be huge and may not fit into the main memory of a single machine.
We therefore study distributed extensions of established single machine clustering algorithms.
This enables us to efficiently compute clusterings in huge graphs.

We consider the problem of clustering a graph into disjoint clusters.
While there is no universally accepted definition of a good clustering, it is commonly accepted that clusters should be internally densely and externally sparsely connected.
Our algorithms optimize two established quality measures that formalize this concept: modularity~\cite{ng-fecsn-04} and map equation~\cite{rab-t-09}.
Other community detection formalizations have been considered.
For example, EgoLP~\cite{bkafktk-elpfd-14} is a distributed algorithm to find overlapping clusters.

\subsection{Related Work}

Existing distributed approaches follow one of two approaches.

The first is to partition the graph into a subgraph per machine.
Each subgraph is then clustered independently on one machine.
Then, all nodes of each cluster are merged summing up weights of parallel edges.
The resulting coarser graph is clustered on a single machine.
This assumes the coarsened graph fits in the memory of a single machine.
In~\cite{zy-a-16}, for the partitioning, the input node ID range is chunked into equally sized parts.
This can work well, but is problematic if input node IDs do not reflect the graph structure.
In~\cite{wfsp-f-14}, the input graph is partitioned using the non-distributed, parallel graph partitioning algorithm ParMETIS~\cite{kk-apamg-98}.
While this is independent of node IDs, it requires that the graph fits into the memory of one machine for the partitioning.

The second approach consists of distributing the clustering algorithm itself.
Using MPI,~\cite{qcpg-scdla-15} have introduced a distributed extension of the Louvain algorithm~\cite{bgll-f-08}.
Similarly,~\cite{lywcl-fcdlw-16} have presented an algorithm that uses the GraphX framework.
Another algorithm named GossipMap is presented in~\cite{bh-gm-15} which uses the GraphLab framework.
Our algorithms also use this second approach.

All of these related algorithms heuristically optimize modularity except GossipMap, which optimizes the map equation.

\subsection{Contribution}

We propose two distributed graph clustering algorithms, \dslmmod{} and \dslmmap{}, that optimize modularity and map equation, respectively.
Our algorithms are the first graph clustering algorithms based on Thrill~\cite{bajlnnssss-thpad-16}, a distributed big data processing framework written in C++ that implements an extended MapReduce model.
Our algorithms are easy to extend for optimizing different density-based quality measures.
To evaluate the clustering quality, we compare against ground truth communities on synthetic LFR~\cite{lfr-b-08} graph clustering benchmark graphs with up to 68 billion edges.
Even for these graphs, 32 hosts of a compute cluster are enough.
Our results show that our algorithms scale well and \dslmmap{} is better at recovering the ground truth than the sequential Infomap algorithm~\cite{rab-t-09}.
On real-world graphs, our algorithms perform similarly to non-distributed baseline algorithms in terms of the quality of the detected clusterings and stability between different runs.
We evaluate both similarities and quality scores, as for quality scores small changes can result in vastly different clusterings~\cite{hmc-p-10}.

Similar to most related work, we make implicit assumptions on the structure of the graph.
We assume that all edges incident to nodes in a single cluster fit into the memory of a single compute node.
In practice, this is only a limitation when a graph has huge clusters.
In many scenarios like social networks or web graphs, this is no limitation as cluster sizes are not that huge.
Our algorithms can be modified to avoid these restrictions, but this would increase running times.

\textbf{Outline.}
In the following we introduce our notation and present the quality measures we optimize.
We also give a brief introduction to Thrill.
In Section~\ref{sec:algorithm}, we present our algorithms.
In Section~\ref{sec:experiments} we present our experimental results.
We conclude in Section~\ref{sec:conclusion}.

\section{Preliminaries}

Conceptually, our algorithms work on undirected graphs.
However, we represent all graphs $G=(V,E, \omega)$ as symmetric, directed, weighted graphs of $|V| = n$ nodes and $|E| = m$ edges.
The pair $(u, v) \in E$ represents the edge from $u$ to $v$.
Unless stated otherwise, there are no multi-edges.
We describe our algorithms for weighted graphs.
As our input graphs are unweighted, we set $\omega(u, v) = 1$ for every edge $(u, v) \in E$.

A cluster $C$ is a node subset.
A clustering $\mathcal{C}$ is a set of clusters such that each node is part of exactly one cluster.

The \emph{weighted degree} $\deg(x)$ of a node $x$ is the sum over the weights of all outgoing edges $(x,y)$ of $x$.
The weight of loop edges is counted twice.
The volume $\vol(C)$ of a set of nodes $C$ is the sum of their weighted degrees.
The cut $\cut(C, D)$ between two sets of nodes $C$, $D$ is the sum of the weights of all edges $(x,y)$ such that $x\in C$ and $y\in D$.
As a simplification, we write $\cut(v, C)$ for $\cut(\{v\}, C)$, $\cut(C) := \cut(C, V\setminus C)$ for the cut between $C$ and the rest of the graph.

Many approaches exist to formalize the quality of a clustering.
In this work, we study two popular ones: modularity~\cite{ng-fecsn-04} and map equation~\cite{rab-t-09}.
The modularity of a clustering $\mathcal{C}$ is defined as
\[
\mathcal{Q}(\mathcal{C}) := \sum_{C \in \mathcal{C}}\frac{\vol(C) - \cut(C)}{\vol(V)} - \sum_{C \in \mathcal{C}}\frac{\vol(C)^2}{\vol(V)^2}\text{.}
\]
Higher modularity values indicate better clusterings.
However, sometimes higher modularity values can also be achieved by merging small but actually clearly distinct clusters.
This effect is called resolution limit~\cite{bf-rlcd-07}.
For the map equation, this effect is much weaker~\cite{kr-e-15}.
Clusterings are better when they have a lower map equation score.
To simplify its definition, we set $\plogp(x):=x \log x$.
The definition is

\begin{align*}
  L(\mathcal{C}) :=& \plogp \left(\sum_{C \in \mathcal{C}}\frac{\cut(C)}{\vol(V)}\right) - 2 \sum_{C \in \mathcal{C}}\plogp \left( \frac{\cut(C)}{\vol(V)}\right)\\
                  & + \sum_{C \in \mathcal{C}}  \plogp \left(\frac{\cut(C) + \vol(C)}{\vol(V)}\right) - \sum_{v \in V} \plogp \left(\frac{\deg(v)}{\vol(V)}\right)\text{.}
\end{align*}
The last term is independent of the clustering and therefore does not affect the optimization.
Thus, we omit it in our algorithms.
Optimizing modularity is NP-hard~\cite{bdgghnw-omc-08} but it can be optimized heuristically in practice~\cite{bgll-f-08}.
Only heuristic map equation optimization algorithms are known to us~\cite{rab-t-09}.

To compare clusterings, either with ground truth communities or with a clustering calculated using a different algorithm, we use
the \emph{adjusted rand index} (ARI)~\cite{ha-c-85}.
The maximum value of 1 indicates that both clusterings are equal.
ARI is normalized such that when one of the two clusterings is random, its expected value is 0.
It can also be negative.

\subsection{Thrill}

Thrill \cite{bajlnnssss-thpad-16} is a distributed C++ big data processing framework.
It can distribute the program execution over multiple machines and threads within a machine.
Each thread is called \emph{worker}.
Every worker executes the same program.

If there is not enough main memory available, Thrill can use local storage such as an SSD as external memory to store parts of the processed data.

Data is maintained in distributed immutable arrays (DIA).
Distributed Thrill operations are applied to the DIAs.
For example, Thrill contains a \emph{sort} operation, whose input is a DIA and whose output is a new sorted DIA.
Similarly, the \emph{zip} operation combines two DIAs of the same length into one DIA where each element is a pair of the two original elements.

Thrill also supports DIAs with elements of non-uniform size, as long as each element fits into the memory of a worker.
This allows elements to be arrays.

Apart from zip and sort, we use the following operations: 
The \emph{map} operation applies a function to each element of a DIA.
The return values are the elements of the new DIA.\@
\emph{Flatmap} is similar, but the function may emit 0, 1, or more elements.
This is similar to the map operation in the original MapReduce model.

A DIA can be \emph{aggregated} by a key component.
All elements with the same key are combined and put into an array.
This is similar to the reduce operation in the original MapReduce model.
An aggregation is much more efficient if the keys are consecutive integers.
In that case, the result is also automatically sorted by the keys.
We use this optimized variant for all aggregations that are based on node IDs.

\section{Algorithm}
\label{sec:algorithm}

The basis of our algorithm is the Louvain algorithm~\cite{bgll-f-08}, a fast algorithm for modularity optimization that delivers high-quality results.
The original Infomap algorithm proposed for optimizing the map equation~\cite{rab-t-09} is based on the same scheme, but introduces additional steps to improve the quality.

Initially, every node is in its own cluster.
This is called a singleton clustering.

In the local moving phase, the Louvain algorithm works in rounds.
In each round, it iterates in a random order over all nodes $v$.
For every $v$, it considers $v$'s current cluster and all clusters $C$ such that there is an edge from $v$ to a node of $C$.
For all these clusters, the difference in quality score $\Delta_{v,C}$ if $v$ was to be moved into $C$ is computed.
If an improvement is possible, $v$ is moved into a cluster with maximal $\Delta_{v,C}$, resolving ties uniformly at random.
The local moving phase stops when no node is moved in a round or a maximum number of rounds is reached.

After the local moving phase, the contraction phase starts.
All nodes inside a cluster are merged into a single node.
The weights of all edges from a cluster $C$ to $D$ are summed up and added as an edge from the node representing $C$ to the node representing $D$.
All edge weights within a cluster are summed up and added as a loop edge.
The contraction does not change the quality score.
On the contracted graph, the algorithm is applied recursively.
It terminates when the contraction phase is entered and the clustering is still the singleton clustering.

\subsection{Distributed Synchronous Local Moving (DSLM)}\label{sec:dslm}

\begin{figure}[tb]
\centering
\begin{tikzpicture}[scale=0.7, every node/.style={transform shape, font=\Large}]
  \node [action] at (0, 0) (input_g) {\node{Input $G$}};
  \node [action, left=.5cm of input_g] (input_c) {\node{Input $\mathcal{C}$}};

  \node [lop, below=2cm of {$(input_g)!0.5!(input_c)$}] (zip)        {\node{Zip}};
  \node [dop, below=1cm of zip]                               (aggregate)  {\node[node:dlm:agg]{Aggregate by $C$}};
  \node [lop, below=1cm of aggregate]                    (flatmap)    {\node[node:dlm:exp]{FlatMap}};
  \node [dop, below=1.7cm of flatmap]                      (aggregate2) {\node[node:dlm:agg_n]{Aggregate by $v$}};
  \node [lop, below=2cm of aggregate2]                   (zip2)     {\node[node:dlm:zip2]{Zip}};
  \node [lop, below=2cm of zip2]                   (select)     {\node[node:dlm:select]{Map}};

  \node [action, below=1.5cm of select]  (output)  {\node{Output}};

  \draw [->] (input_g) -- (zip) node [midway,right=.3cm] () [align=center] {$(v,\langle u_i \rangle, \langle w_i \rangle)$ :\\sorted by $v$};
  \draw [->] (input_c) -- (zip) node [midway, left] (foo) [align=center] {$(v, C)$ :\\sorted by $v$};

  \draw [->] (zip) -- (aggregate) node [midway,left] {$(C, v, \langle u_i \rangle, \langle w_i \rangle)$};
  \draw [->] (aggregate) -- (flatmap) node [midway,left] {$(C, \langle v_i, \langle u_j \rangle, \langle w_j \rangle \rangle)$};
  \draw [->] (flatmap) -- (aggregate2) node [midway,right] () {$(C, v, \vol(C\setminus{v}), \cut(v, C\setminus{v}), \cut(C\setminus{v}))$};
  \draw [->] (aggregate2) -- (zip2) node [midway,right] (foo) [align=left] {sorted by $v$ : \\$(v, \langle C_i, \vol(C_i\setminus{v}),$ $\cut(v, C_i\setminus{v}), \cut(C_i\setminus{v}) \rangle)$};
  \draw [->] (zip2) -- (select) node [midway,right] (foo) [align=left] {sorted by $v$ : \\$(v, \langle u_i \rangle, \langle w_i \rangle, \langle C_i, \vol(C_i\setminus{v}), \cut(v, C_i\setminus{v}), \cut(C_i\setminus{v}) \rangle)$};
  \draw [->] (select) -- (output) node [midway,right] {sorted by $v$ : $(v, C^*)$};

  \draw [->] (output.west) -- ++(-3,0) |- (input_c.west);
  \draw [->] (input_g.east) -- ++(7,0) |- (zip2.east);

  \node [code, right=1cm of flatmap] (code_flatmap) [align=left] {%
    Calculate and emit cluster\\ data for active neighbors
  };
  \draw [dotted] (code_flatmap) -- (flatmap);

  \node [code]  at (code_flatmap |- select) (code_select) {%
    Select best cluster
  };
  \draw [dotted] (code_select) -- (select);

  \node [code]  at (code_flatmap |- output) (explanation_out) [align=left] {%
    Combined with round input \\ for clusters of inactive nodes
  };
  \draw [dotted] (explanation_out) -- (output);
\end{tikzpicture}
\caption{DSLM data flow}\label{fig:dslm_data_flow}
\end{figure}
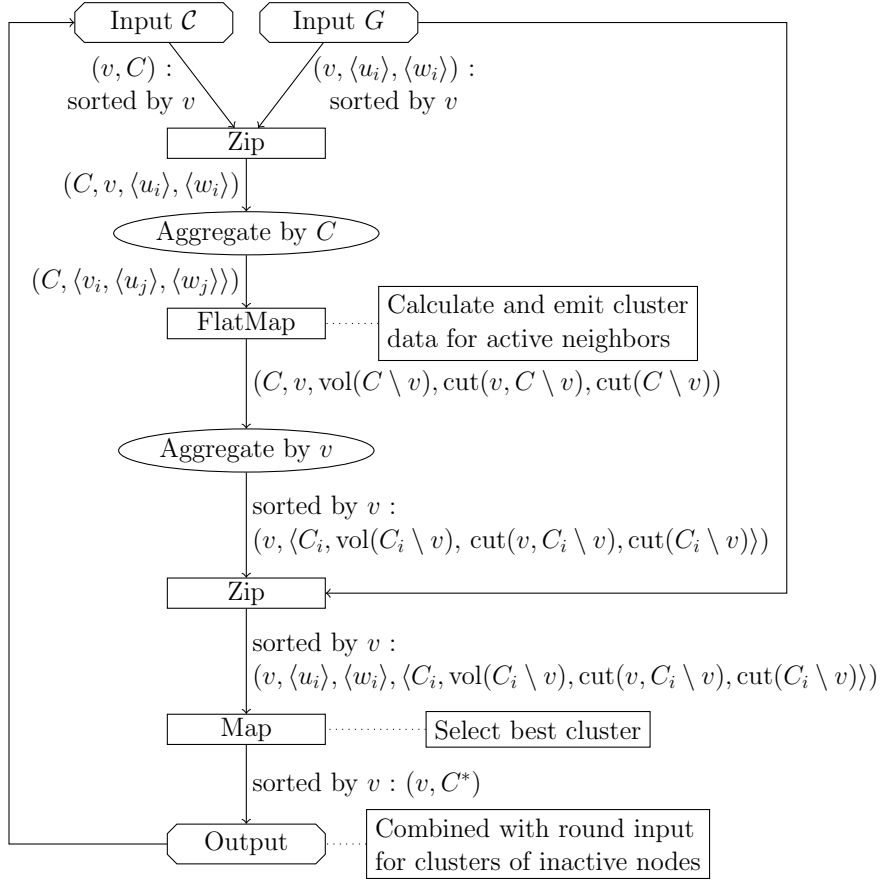

In local moving, the $i$-th move depends on the clustering after the first $i-1$ moves are executed.
This data dependency makes the parallelization difficult.
We therefore split a round into sub-rounds.
Every node picks a random sub-round in which it is \emph{active}.
In the $i$-th sub-round, all active nodes are moved synchronously and in parallel with respect to the clustering after the $(i-1)$-th sub-round.
For the first sub-round, this is with respect to the initial clustering.
We call this scheme \emph{synchronous local moving}. 
For our \emph{distributed} synchronous local moving,
a global hash function $h$ maps a tuple of a node ID $v$, the number of completed rounds and a global seed onto the sub-round $r_v$ in which $v$ is active.
Figure~\ref{fig:dslm_data_flow} illustrates the data flow of our algorithm.

We represent a graph and its clustering as two DIAs.
They have length $n$ and are stored sorted by their first item, the node ID $v$.
The graph DIA stores triples $(v,\langle u_i \rangle, \langle w_i \rangle)$, where $\langle u_i \rangle$ and $\langle w_i \rangle$ are equally-sized arrays.
For every $i$, there is an edge $(v,u_i)$ with weight $w_i$.
The clustering DIA of pairs $(v,C)$ stores for every node $v$ its cluster ID $C$.

In DSLM, a sub-round is composed of a bidding and a compare step.
In the bidding step, the clusters place bids for active nodes.
In the compare step, every active node compares its bids and becomes part of the cluster with the best bid.

To allow a node $v$ to compute the change in modularity or map equation when joining the neighboring cluster $C$, each bid contains:
(a) volume $\vol(C\setminus{v})$,
(b) cut weight $\cut(v, C\setminus{v})$ between $C\setminus{v}$ and $v$, and
(c) cut weight $\cut(C)$ between $C$ and the remaining graph.

The bidding step starts by zipping the clustering DIA and graph DIA and aggregating this zipped DIA by cluster ID.\@
The result is a DIA with one large element per cluster $C$.
Each element contains all nodes in the cluster $C$ and the neighborhoods of these nodes.
This allows us to compute the measures (a), (b), and (c) using a non-distributed algorithm inside a worker.
Using a flatmap, our algorithm emits for every cluster $C$ bids for all active nodes inside $C$ and adjacent to $C$.
It can determine which nodes are active as the hash function $h$ is globally known.
The generated bid DIA consists of quintuples $(C, v, \vol(C\setminus{v}),$ $\cut(v, C\setminus{v}), \cut(C\setminus{v}))$.
Each quintuple is a bid of cluster $C$ for active node $v$.

The compare step aggregates the bid DIA by node $v$.
After the aggregation, the result is zipped with the graph DIA to obtain the nodes' degree and loop edge weight.
In a map operation, we use this information to compute for every active node the best bid and return the according cluster.
We retrieve the old cluster ID for non-active nodes by zipping with the original clustering DIA.
This yields the updated clustering DIA, which
is the input of the next sub-round.

\textbf{Implementation Details and Optimizations.}
If modularity is optimized, our algorithm can be improved in two ways.
First, we can omit $\cut(C\setminus{v})$ as it is only needed for the map equation.
Second, we can compare bids without knowing the current cluster.
This allows us to use a pairwise reduction instead of one that first waits for all elements.
As we still need the node's degree, each worker stores the degree of the nodes that are reduced on that worker in a plain array.
This is possible because we know on which worker each node will end up.

\subsection{Distributed Contraction and Unpacking}

The contraction is performed in three steps:
(a) obtain consecutive cluster IDs,
(b) replace all node IDs by the cluster IDs, and
(c) combine multi-edges.

We first zip the graph and clustering DIAs and aggregate them by cluster ID.
To get consecutive cluster IDs, we replace them with the element positions.
From the result, which contains tuples $(C, \langle v_i, \langle u_j \rangle, \langle w_j \rangle \rangle)$, we derive two DIAs.

The first DIA is a new clustering DIA with consecutive cluster IDs.
To obtain it, we first drop the neighborhood information.
We store this intermediate DIA $(C, \langle v_i \rangle)$ for the unpacking phase.
Then, we expand it into pairs $(v_i, C)$ of node and cluster ID using a flatmap operation and sort by node IDs.

The second DIA is the contracted graph DIA.
To obtain it, we first emit a triple $(C_u, v, w)$ for every node $v$ that is a neighbor of a node in $C_u$ in a flatmap operation.
The weight $w$ is the sum of all edge weights from nodes in $C_u$ to $v$.
We aggregate this DIA by $v$, zip it with the new clustering DIA and replace $v$ by $v$'s cluster ID $C_v$.
We then aggregate by $C_v$ to obtain pairs $(C_v, \langle C_{u,i}, w_{i} \rangle )$ containing the neighboring clusters $C_{u,i}$ for every $C_v$.
Finally, we sum up the weights $w_i$ for the same neighboring cluster for every cluster $C_v$ in a map operation.

To unpack the clustering calculated in a level, we zip the clustering DIA $(v, C_v)$ with the intermediate clustering DIA $(v,\langle v_i \rangle)$ of a cluster $v$ and its nodes $\langle v_i \rangle$ from the previous contraction phase.
A flatmap operation assigns the cluster ID $C_v$ of the contracted node to all original nodes $u \in \langle v_i \rangle$, resulting in a clustering DIA $(u, C_u)$.
After sorting it by node, it is returned to the next level.

\section{Experiments}
\label{sec:experiments}

In this section, we present an experimental evaluation of our algorithm DSLM\footnote{This paper only covers parts of our experiments. Under \url{https://github.com/kit-algo/distributed\_clustering\_thrill\_evaluation} you can find additional analyses, links to our raw data and information on how to explore our data on your own.}.
The source code of our implementation is publicly available on GitHub\footnote{\url{https://github.com/kit-algo/distributed\_clustering\_thrill}}.
We first describe our experimental setup.
Then, we present weak scaling experiments to evaluate the running time, compare the quality on LFR benchmark graphs~\cite{lfr-b-08} and evaluate the performance on established real-world benchmark data.

All running time experiments were performed on a compute cluster.
Each compute node has two 14-core Intel Xeon E5-2660 v4 processors (Broadwell) with a default frequency of 2~GHz, 128~GiB RAM and 480~GiB SSD.
They are connected by an InfiniBand 4X FDR Interconnect.
We use the TCP back-end of Thrill due to problems with the combination of multithreading and OpenMPI.
We use Thrill's default parameters, except for the block size, which determines the size of data packages sent between the hosts.
Preliminary experiments found that a block size of 128~KiB instead of the default 2~MiB yields the best results.

For our algorithms, we use four sub-rounds as suggested in a preliminary study~\cite{z-edgcm-17}.
Using less results in problems with the convergence.
Using more does not significantly improve quality but increases running time.
In each local moving phase, we perform at most eight rounds.
All experiments were performed with 16 threads per host.
More threads do not improve the running times much further.
Preliminary experiments indicate that the performance is RAM bound.

Apart from \dslmmod{} and \dslmmap{} that optimize modularity and map equation, we also evaluate a variant \dslmmodwocont{} that stops after the first local moving phase.
This significantly decreases the running time and surprisingly also improves the quality on synthetic instances.
We evaluate this behavior in more detail in Section~\ref{sec:quality}.

For modularity, we compare against our own implementation of the sequential Louvain algorithm~\cite{bgll-f-08} and the shared-memory parallel PLM~\cite{sm-epacd-16}.
For map equation, we compare against the sequential Infomap~\cite{rab-t-09}, the shared-memory parallel RelaxMap~\cite{bhwrh-sefbc-17} and the distributed GossipMap~\cite{bh-gm-15} implementations.

In a preprocessing step, we remove degree zero nodes, make the ID space consecutive and randomize the node order.
This ensures that our algorithms are independent of input order and improves load balancing.

All experiments were repeated 10 times with different random seeds.
We report averaged results and standard deviation where possible as error bars.

During the experiments, the meltdown and spectre vulnerabilities became public and performance impacting patches were applied to the machines.
Rerunning some experiments showed a slowdown of up to 1.6 for runs with 32 hosts but no significant slowdown for runs with a single host.
We did not have the resources to rerun all experiments.
Also, we expect the performance of the machines to change further in the future.
Patches with less impact (Retpoline) are available but have not been rolled out yet.
More vulnerabilities have been discovered in the meantime and it is unclear if fixes for them will have further performance implications\footnote{\url{https://securityaffairs.co/wordpress/72158/hacking/spectre-ng-vulnerabilities.html}}.
At the point of initial patch distribution, most distributed algorithm runs were already done.
About half of the GossipMap runs on the real world graphs were performed afterwards and are excluded from the running time reports.
All runs for non-distributed algorithms were performed with patches applied, as their performance should not have been affected significantly.

\textbf{Synthetic Instance Generation.}
Our synthetic test data is generated using the established LFR benchmark generation scheme~\cite{lfr-b-08}.
To generate graphs of up to 512 million nodes and 67.6 billion (undirected) edges in a reasonable time, we use the external memory LFR generator implementation of~\cite{hmpw-iogmg-17}.

LFR benchmark graphs feature a ground truth clustering.
Node degrees and cluster sizes are drawn from power law distributions.
The mixing parameter $\mu$ determines the fraction of edges that are between different clusters.
For details, we refer the reader to the original description~\cite{lfr-b-08}.
We set a minimum degree of 50 and a maximum degree of 10\,000 with a power law exponent of 2.
This leads to an average degree of approximately 264.
For the communities, we set 50 as minimum and 12\,000 as maximum size with a power law exponent of 1.
Unless otherwise noted, we set the mixing parameter $\mu$ to 0.4.

\subsection{Weak Scaling}

\begin{figure}[tb]
  \includegraphics[width=0.48\textwidth]{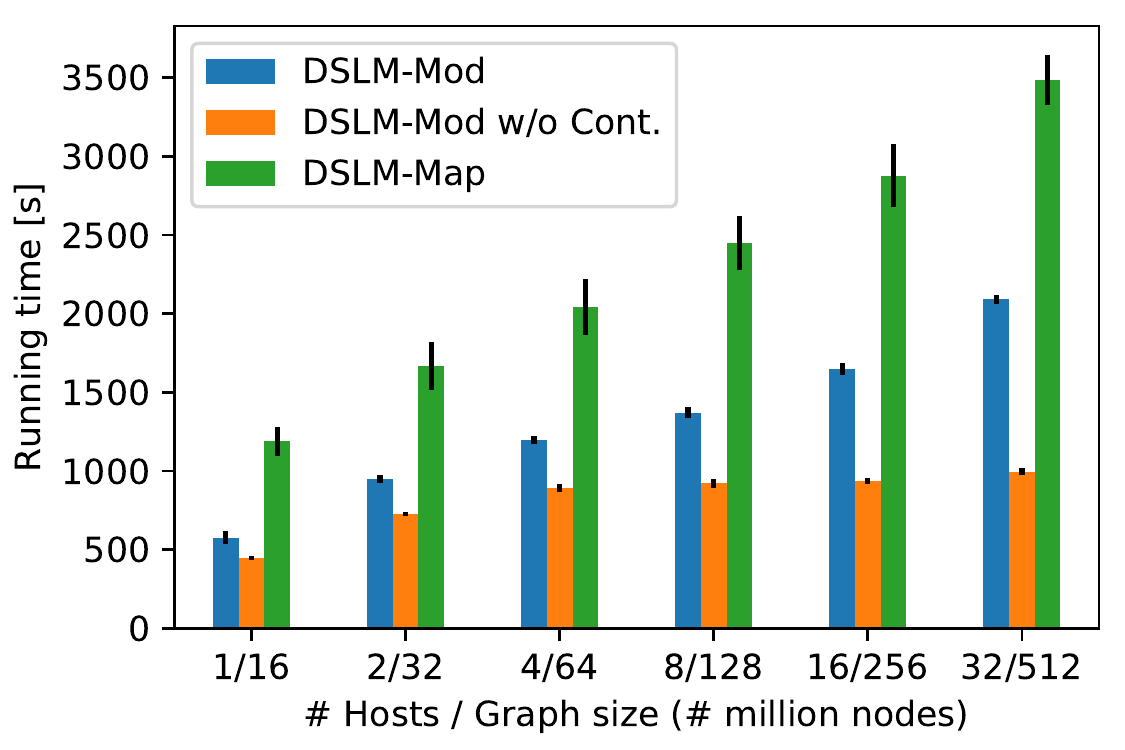}
  \includegraphics[width=0.48\textwidth]{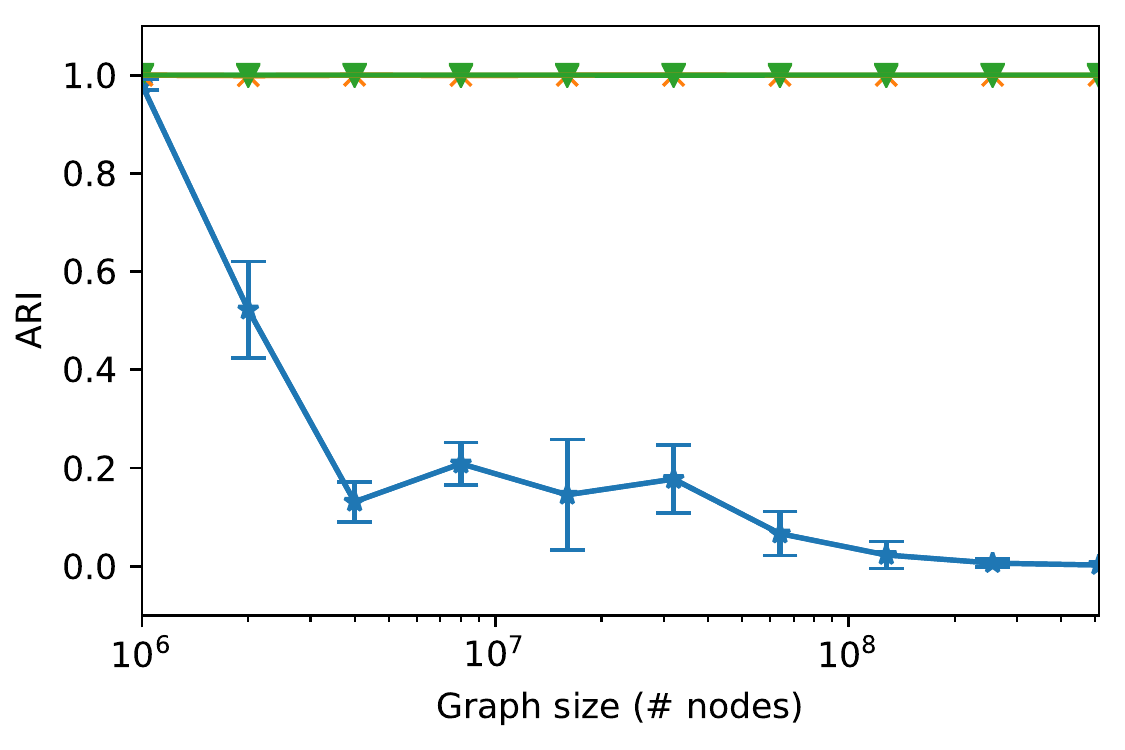}
\caption{Weak scaling: running time of our distributed algorithms and ARI with ground truth. The \dslmmodwocont{} ARI line is hidden by the \dslmmap{} line.}\label{fig:weak_scaling}
\end{figure}

For the weak scaling experiments, we use LFR graphs with 16, 32, 64, 128, 256 and 512 million nodes.
We cluster them on 1, 2, 4, 8, 16, and 32 hosts respectively.
The left part of Figure~\ref{fig:weak_scaling} shows the running time of our algorithms.
Our algorithms utilize almost the entire available RAM.
GossipMap is less memory-efficient and was unable to cluster the graphs in these configurations and crashed.

With a linear time algorithm and perfect scaling, we would expect that the running time remains constant as we increase graph size and the number of nodes.
For the variant of \dslmmodwocont{}, the running time actually does not increase much.
The running time of the full \dslmmod{} and \dslmmap{} algorithms increases approximately linearly though as the number of hosts is scaled exponentially.
The reason for this is that LFR graphs get very dense during contraction and thus in particular larger graphs still have a significant amount of edges after the contraction.
Also, \dslmmap{} is approximately a factor of two slower than \dslmmod{}.
This is expected as the optimizations described at the end of Section~\ref{sec:dslm} are not applicable to \dslmmap{}.

\subsection{Quality}\label{sec:quality}

First, we evaluate the quality of the clusterings obtained in the weak scaling experiment.
The right part of Figure~\ref{fig:weak_scaling} depicts the similarities of the clusterings found by our algorithms and the ground truth.
From the plot, we observe that \dslmmap{} finds a clustering very close to the ground truth.
\dslmmodwocont{} achieves similar results.
Unfortunately, \dslmmod{} fails to find a clustering similar to the ground truth on the larger instances.
This shows that after the contraction, clusters are merged that should not be merged.
To verify if the worse results of \dslmmod{} are due to the resolution limit, we started a sequential Louvain algorithm on a graph where we contracted the ground truth.
This algorithm indeed merges clusters, showing that the resolution limit is relevant here.
However, the thereby detected clusters are much more similar to the ground truth than those detected by \dslmmod{} and even the ground truth alone has higher modularity scores than those found by \dslmmod{}.

\begin{figure*}[tb]
  \centering
  \includegraphics[width=0.98\textwidth]{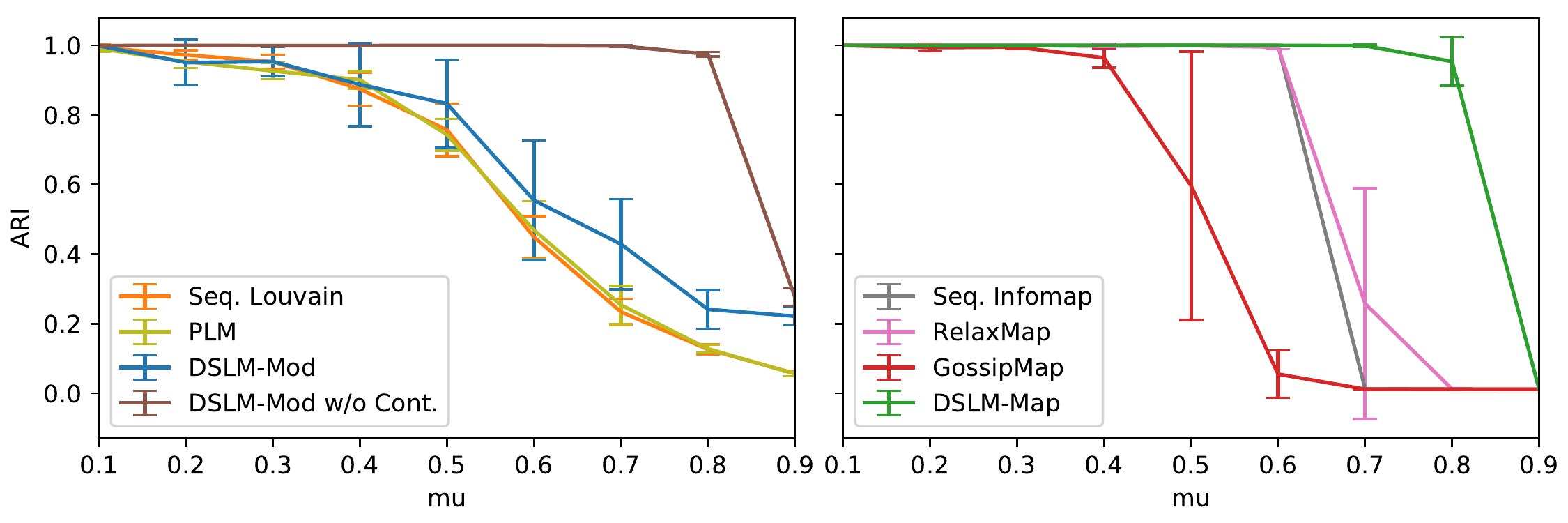}
  \caption{Adjusted rand index with ground truth for $\mu \in [0.1, 0.9]$.}
  \label{fig:quality}
\end{figure*}

We also use smaller LFR graphs with 1M nodes and varying mixing parameter to compare the quality of the communities found by all compared algorithms.
Figure~\ref{fig:quality} shows the adjusted rand index of the detected clusterings with the ground truth.
\dslmmodwocont{} and \dslmmap{} outperform all other algorithms by a significant margin.
On average, \dslmmod{} still outperforms the other modularity-optimizing algorithms.
For all values of $\mu$, the ground truth has a higher modularity score than the clustering found by the modularity-optimizing algorithms.
Merging clusters of the ground truth again improves the modularity score but leads to clusterings that still have an ARI of above $0.99$ for $\mu < 0.9$.
With the algorithms optimizing map equation, the situation is similar.
For $\mu < 0.8$, the ground truth, which \dslmmap{} consistently finds, has a better map equation score than the clusterings found by all other algorithms.
For $\mu \ge 0.8$, a singleton clustering yields a better map equation score than the ground truth clustering.
GossipMap finds neither good map equation scores nor the ground truth for $\mu > 0.4$.

Overall, for these LFR benchmark graphs, DSLM seems to be superior to sequential local moving.
Examining sequential local moving algorithms, we noticed that high-degree nodes attract many nodes in the first local moving round.
After a few nodes join their cluster, many others follow.
In contrast to that, with DLSM, 25\% of the nodes can join the cluster of another node before any cluster sizes come into play.
Apparently, this avoids such accumulation effects.

\subsection{Real-World Graphs}

To assess whether our results on LFR benchmark graphs are also true for real-world graphs, we performed experiments on a set of different real-world graphs.
From the Stanford Large Network Dataset Collection~\cite{lk-snapd-14}, we include
three social networks (com-LiveJournal, com-Orkut and com-Friendster). 
From the 10th DIMACS Implementation challenge~\cite{bmsskw-b-13}, we include two web graphs where nodes represent URLs and edges links between them (uk-2002 and uk-2007-05). 

\begin{table*}[tb]
  \centering
  \caption{Average running time in seconds of the algorithms on the real-world graphs.}
  \label{tab:real_world_runtime}
  \input{plots/real_world_runtimes.tex}
\end{table*}

We clustered these graphs both with the sequential baseline algorithms and our distributed algorithms.
Table~\ref{tab:real_world_runtime} depicts the sizes of the graphs, the number of hosts we used for the distributed algorithms and the running times in seconds.
RelaxMap and GossipMap use the directed version of the map equation that includes a PageRank approximation as preprocessing step.
To allow for a fair comparison, we only report the running time of the actual clustering step after this preprocessing.
As three GossipMap runs on uk-2007-05 crashed, there are less samples.
Unfortunately, both the original Infomap implementation and RelaxMap were not able to cluster all instances.
On the two largest graphs, 128~GB of RAM were not enough memory (oom).

With 8 or 16 hosts, our distributed algorithms are almost always faster than the sequential and shared-memory parallel algorithms.
Note that due to the randomized node order, PLM is slower in our experiments than reported in~\cite{sm-epacd-16}.
\dslmmap{} with 8 hosts is more than a factor of 5, for uk-2002 even a factor of 20 faster than RelaxMap and also a factor of 10 faster than GossipMap.
\dslmmod{} is faster than \dslmmap{}, but the difference is less pronounced than in the weak scaling experiments.
This shows the advantage of our algorithmic scheme in combination with the efficient Thrill framework.

\begin{table*}[tb]
  \caption{Average modularity/undirected map equation scores obtained by the respective algorithms.}
  \label{tab:real_world_quality_scores}
  \centering
  \input{plots/quality_scores_table.tex}
\end{table*}

Table~\ref{tab:real_world_quality_scores} shows the average modularity and map equation scores obtained by the algorithms.
We observe that PLM on average finds the best modularity scores with a minor exception on uk-2002 where Louvain finds better values.
\dslmmod{} performs slightly worse, \dslmmodwocont{} significantly worse.
This shows that \dslmmodwocont{}, which performed really well on the LFR benchmark graphs, is unsuited for real-world graphs.

The best map equation scores are found by the Infomap algorithm where it finished the computation.
Since RelaxMap and GossipMap use the directed map equation, we also include the directed Infomap algorithm to evaluate if using the directed map equations leads to different results.
Surprisingly, in some cases Infomap optimizing the directed map equation finds better clusterings with respect to the undirected map equation than the undirected Infomap, though the differences are very small.
On the two smallest graphs, RelaxMap finds better clusterings than the distributed algorithms.
On uk-2002, RelaxMap is outperformed by \dslmmap{}.
\dslmmap{} finds better clusterings than GossipMap on all graphs except for LiveJournal, on the two largest graphs by a significant margin.
Since modularity and map equation feature counterintuitive behavior like the resolution limit, quality scores on their own can be misleading.
We therefore also compare the obtained clusterings in terms of ARI.

\begin{table*}[tb]
  \caption{Average similarities in terms of ARI with best clustering found according to the respective quality score. Underlined entries indicate the algorithm which found the clustering with the best score.}
  \label{tab:real_world_quality}
  \centering
  \input{plots/quality_table.tex}
\end{table*}

Among all algorithms that optimize the same quality score, we determine for each graph the detected clustering with best score.
We use these best clusterings as baselines to which we compare all other detected clusterings that were detected optimizing the same quality measure.
Table~\ref{tab:real_world_quality} shows the average similarity in terms of adjusted rand index and highlights which algorithm detected the used baseline clustering.
In most cases, this is the sequential baseline.

For modularity, this is in contrast to the results from Table~\ref{tab:real_world_quality_scores} where on average, PLM outperforms Louvain for most graphs.
Only on Friendster and uk-2002, the modularity-optimizing algorithm that found the best clustering also has the highest average quality scores.
We observe that the modularity-optimizing algorithms do not consistently find the same clustering.
Clusterings may vary vastly depending on the random seed.
Further, on social networks the adjusted rand indices are smaller than on web graphs.
This is probably due to web graphs having a more pronounced community structure.
\dslmmod{} produces clusterings that are less similar, but still much more similar than \dslmmodwocont{}, which produces vastly different clusterings.
This confirms our observation from Table~\ref{tab:real_world_quality_scores}.
Omitting the contraction significantly decreases the quality of clusterings on real world graphs.

Infomap is in general much more stable than Louvain with an adjusted rand index close to 1.
\dslmmap{} produces very similar clusterings on uk-2002.
On the LiveJournal and Orkut graphs, the clusterings are slightly less similar.
Quite interestingly, the parallel RelaxMap and the distributed GossipMap algorithms produce significantly different clusterings in particular for the two social networks.
As the results of the directed Infomap algorithm shows, this is not due to optimizing the directed map equation.
We conclude that RelaxMap and GossipMap indeed fail to find similar clusterings reliably.

\section{Conclusion}
\label{sec:conclusion}

We have introduced two distributed graph clustering algorithms, \dslmmod{} and \dslmmap{}, that optimize modularity and map equation, respectively.
They are based on the Thrill framework. 
In an extensive experimental evaluation, we have shown that on LFR benchmark graphs, \dslmmap{} achieves excellent results, even better than the sequential Infomap algorithm.
For \dslmmod{}, we also evaluate a variant without contraction which has great performance on LFR benchmark graphs.
The full \dslmmod{} algorithm with contraction fails to recover the ground truth on LFR benchmark graphs -- similar to the sequential Louvain algorithm -- but significantly outperforms the variant without contraction on real-world graphs.
On real-world graphs, both distributed algorithms find clusterings only slightly different than the sequential algorithms.
Compared to GossipMap, the state-of-the-art distributed algorithm for optimizing map equation, \dslmmap{} is up to an order of magnitude faster while detecting clusterings that have similar or better map equation scores and are more similar to the clustering with the best map equation score.

In the first local moving phase, synchronous local moving seems to be superior to sequential local moving.
Further research is needed to see if this is a phenomenon particular to the LFR graphs we studied or if synchronous local moving could be a way to avoid local maxima when optimizing such quality functions.
After the contraction, more careful local moving strategies should be developed though to avoid the problems we see in particular on LFR graphs.
Therefore, further research on different local moving strategies seems to be a promising direction.

\input{main.bbl}
\end{document}

%% file: plots/real_world_runtimes.tex
{\setlength{\tabcolsep}{0.38em}
\begin{tabular}{lrrrrrrrrrrr}
\toprule
{} & \rot{\# Nodes} & \rot{\# Edges} & \rot{\# Hosts} & \rot{Louvain} & \rot{PLM} & \rot{\dslmmod} & \rot{\parbox{1.7cm}{\dslmmodwocont}} & \rot{Infomap} & \rot{RelaxMap} & \rot{GossipMap} & \rot{\dslmmap} \\
\midrule
LiveJournal             &         4M &           34M &      8 &          99 &    25 &      31 &                      14 &        1329 &     163 &      372 &      49 \\
Orkut          &         3M &          117M &      8 &         170 &    53 &      47 &                      34 &        2405 &     415 &      700 &      84 \\
uk-2002        &        18M &          261M &      8 &         572 &   142 &      46 &                      22 &        6656 &     240 &      682 &      52 \\
Friendster     &        66M &         1806M &     16 &        6002 &  1755 &    1047 &                     742 &         oom &     oom &    13743 &    1161 \\
uk-2007-05     &       105M &         3302M &     16 &        7993 &  2520 &     151 &                     106 &         oom &     oom &     4211 &     214 \\
\bottomrule
\end{tabular}
}

%% file: plots/quality_scores_table.tex
{\setlength{\tabcolsep}{0.38em}
\begin{tabular}{lrrrr|rrrrr}
\toprule
{}          & \rot{Louvain} & \rot{PLM} & \rot{\dslmmod} & \rot{\parbox{1.7cm}{\dslmmodwocont}} & \rot{Infomap} & \rot{\parbox{1.5cm}{Directed Infomap}} & \rot{RelaxMap} & \rot{GossipMap} & \rot{\dslmmap} \\
\midrule
LiveJournal &         0.752  & \textbf{0.752} &     0.749 &                     0.591 & \textbf{9.899} &                  9.900  &     9.943 &      9.963 &          9.981 \\
Orkut       &         0.664  & \textbf{0.666} &     0.658 &                     0.524 &        11.826  &         \textbf{11.825} &    11.849 &     11.979 &         11.896 \\
uk-2002     & \textbf{0.990} &         0.990  &     0.990 &                     0.879 &         6.458  &          \textbf{6.458} &     6.476 &      6.550 &          6.468 \\
Friendster  &         0.622  & \textbf{0.627} &     0.616 &                     0.553 &           oom  &                    oom  &       oom &     16.271 & \textbf{14.785} \\
uk-2007-05  &         0.996  & \textbf{0.996} &     0.996 &                     0.919 &           oom  &                    oom  &       oom &      9.034 &  \textbf{8.057} \\
\bottomrule
\end{tabular}
}

%% file: plots/quality_table.tex
{\setlength{\tabcolsep}{0.38em}
\begin{tabular}{lrrrr|rrrrr}
\toprule
{}          & \rot{Louvain} & \rot{PLM} & \rot{\dslmmod} & \rot{\parbox{1.7cm}{\dslmmodwocont}} & \rot{Infomap} & \rot{\parbox{1.5cm}{Directed Infomap}} & \rot{RelaxMap} & \rot{GossipMap} & \rot{\dslmmap} \\
\midrule
LiveJournal & \underline{0.571} &            0.639  &     0.600 &                     0.179 &            0.976  &          \underline{0.973} &     0.376 &      0.784 &            0.769  \\
Orkut       & \underline{0.632} &            0.625  &     0.659 &                     0.220 & \underline{0.919} &                     0.925  &     0.807 &      0.491 &            0.819  \\
uk-2002     & \underline{0.730} &            0.724  &     0.674 &                     0.047 &            0.986  &          \underline{0.985} &     0.928 &      0.698 &            0.970  \\
Friendster  &            0.640  & \underline{0.623} &     0.569 &                     0.361 &              oom  &                       oom  &       oom &      0.013 & \underline{0.748} \\
uk-2007-05  & \underline{0.873} &            0.877  &     0.816 &                     0.279 &              oom  &                       oom  &       oom &      0.132 & \underline{0.986} \\
\bottomrule
\end{tabular}
}

